\documentclass[11pt,a4paper]{article}
\usepackage{times,latexsym}
\usepackage{url}
\usepackage[edges]{forest} 
\usepackage{float}
\usepackage{soul}
\usepackage{graphicx}
\usepackage{adjustbox}
\usepackage{tikz}
\usetikzlibrary{shapes, positioning}
\usepackage{enumitem}
\usetikzlibrary{positioning, arrows.meta, calc, decorations.pathreplacing}
\usepackage[T1]{fontenc}
\usepackage{amsmath}
\usepackage{amssymb}
\usepackage{titlesec}
\usepackage{ACL2023}
\usepackage{natbib}

\title{A Survey on Evaluation Metrics for Music Generation}

\author{ Faria Binte Kader
  \\
  University of Central Florida
  \\
  \texttt{fariabinte.kader@ucf.edu} \\\And
  Santu Karmaker
  \\
  University of Central Florida
  \\
  \texttt{santu@ucf.edu} \\}
\author{
  Faria Binte Kader
  \\
  University of Central Florida
  \\
  \texttt{fariabinte.kader@ucf.edu}
  \And
  Santu Karmaker
  \\
  University of Central Florida
  \\
  \texttt{santu@ucf.edu}
}

\date{}

\begin{document}
\maketitle
\begin{abstract}
Despite significant advancements in music generation systems, the methodologies for evaluating generated music have not progressed as expected due to the complex nature of music, with aspects such as structure, coherence, creativity, and emotional expressiveness. In this paper, we shed light on this research gap, introducing a detailed taxonomy for evaluation metrics for both audio and symbolic music representations. We include a critical review identifying major limitations in current evaluation methodologies which includes poor correlation between objective metrics and human perception, cross-cultural bias, and lack of standardization that hinders cross-model comparisons. Addressing these gaps, we further propose future research directions towards building a comprehensive evaluation framework for music generation evaluation.
\end{abstract}

\section{Introduction}
Recent advancements in computational music research have significantly improved the ability of machines to understand and generate music \citep{yuanchatmusician, copet2024simple, schneider2024mousai}. Large Language models \cite{chang2024survey} and Diffusion-based models \cite{yang2023diffusion} have now the ability to compose and edit melodies, even generate complex musical pieces that mimic human creativity \citep{yu2023musicagent, zhang2023loop}. 
One such example is Suno.ai\footnote{\url{https://suno.com/}}, a web-based service that, given a simple prompt with lyrics, can generate a full song, adding a singing voice within seconds. 
While generative models continue to improve, music generation evaluation at a large scale still lacks standardized assessment criteria due to the inherently subjective and multidimensional nature of musical quality.

A wide range of evaluation metrics has been proposed to assess the quality of both generated audio and symbolic music scores, from statistical comparisons~\cite{chen2024sympac} to machine learning-based similarity measures~\cite{suzuki2023comparative} between generated and reference music. Some metrics also focus on specific musical features, such as melody \cite{yu2022museformer}, rhythm \cite{sheng2021songmass}, harmony \cite{harte2006detecting}, and emotional expression \cite{imasato2023using}. In addition, human evaluation is used to rate subjective qualities like overall quality and prompt alignment, which remain essential for judging expressiveness and creativity.

Unfortunately, as we discuss in detail later in the paper, these metrics rarely capture the complexities of human musical perception. The challenge lies in balancing quantitative measures with subjective listening studies \cite{yang2020evaluation}, as musical quality is often tied to aesthetic preference, cultural background, and contextual interpretation \cite{huron2001music}. While benchmarks such as MARBLE~\cite{yuan2023marble} and MusicTheoryBench ~\cite{yuanchatmusician} offer standard evaluation methods for music understanding and retrieval tasks, no comprehensive framework exists for evaluating generated music scores. To highlight the gravity of this significant gap in the current literature, we provide, in this paper, a comprehensive overview of the evaluation metrics currently used in music generation tasks. We examine computational evaluation techniques, highlighting current limitations, and propose a direction for future improvements. By analyzing existing evaluation strategies, this work aims to shed light on ongoing efforts to develop more robust, interpretable, and standardized music evaluation frameworks.

\section{Background on Computational Music}
\subsection{Music Representation}
Existing music representation techniques deal with two types of music data- audio and symbolic scores to make them computer-interpretable. 

\smallskip
\noindent \textbf{Audio Representations }like Log Mel Spectrograms \cite{logan2000mel}, MFCCs \cite{davis1980comparison}, and Chroma Features \cite{takuya1999realtime} transform raw audio waveforms into machine usable formats for generation and analysis tasks. Pre-trained text-audio encoders like CLAP \cite{elizalde2023clap} and MuLan \cite{huang2022mulan} jointly represent audio and text in the same embedding space.

\smallskip
\noindent \textbf{Symbolic Representations} like MIDI \cite{rothstein1995midi}, MusicXML \cite{goodmusicxml}, ABC Notation \cite{walshaw2021abc}, LilyPond \cite{nienhuys2003lilypond} etc. represent pitch, rhythm, and dynamics in text or event-based form and are widely used in generating and editing text-based musical scores. To make symbolic music more suitable for machine learning, various tokenization methods such as REMI \cite{huang2020pop}, SMT-ABC \cite{qu2024mupt}, Octuple \cite{zeng2021musicbert} etc. encode attributes like pitch, duration and timing data into sequences of tokens.

\subsection{Music Generation Models}
A big part of music computational research is Music Generation. Based on the representations, recent advancements in music generation models can be categorized into two variations-

\smallskip
\noindent \textbf{Audio Music Generation Models }made sequential advancements from transformer-based models like MusicLM \cite{agostinelli2023musiclm} and MusicGen \cite{copet2024simple} to diffusion-based models like Noise2Music \cite{huang2023noise2music}, Mo$\hat{u}$sai \cite{schneider2024mousai}, AudioLDM2 \cite{liu2024audioldm} and ERNIE-Music \cite{zhu2023ernie}. These models can generate good quality music from textual descriptions. Recent advancements in music generation include commercial websites like Suno\footnote{\url{https://suno.com/}} along with open-source models- Yue \cite{yuan2025yue}, SongGen \cite{liu2025songgen}, Ace-Step \cite{gong2025ace} and DiffRhythm \cite{ning2025diffrhythm} that can generate full length songs with proper voice coordinated lyrics.

\smallskip
\noindent \textbf{Symbolic Music Generation Models} focus on producing musical scores in formats like MIDI or ABC notation and are capable of generating multi-instrument compositions. Unfortunately due to the textual nature of the representations, these models can not produce realistic vocals. Symbolic Music Generation is particularly useful for music composing, understanding and editing. The symbolic generation models underwent significant improvements as well from utilizing GANs (MuseGAN \cite{dong2018musegan}) and transformers (Museformer \cite{yu2022museformer}) to diffusion-based models like SD-Muse \cite{zhang2023sdmuse}. 

\begin{figure*}[!htb]  
    \centering
    \includegraphics[width=0.9\textwidth, keepaspectratio]{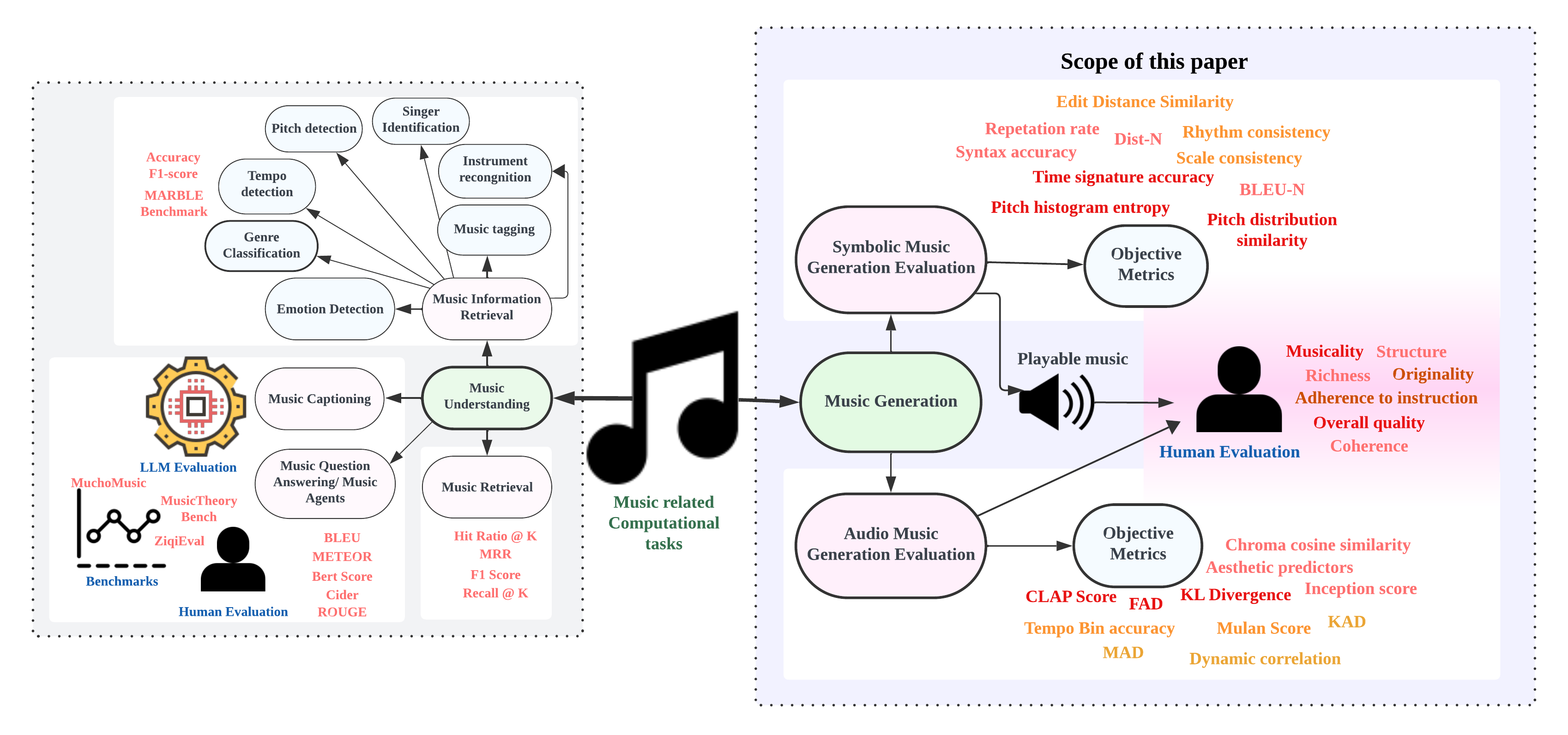}  
     \vspace{-3mm}
     \caption{An illustration of music-related tasks with their corresponding evaluation metrics. Unlike other tasks, music generation evaluation lacks standardized metrics, which is the focus of this survey paper.
     }
    \label{fig:ev}
    \vspace{-2mm}
\end{figure*}

\subsection{Datasets and Benchmarks}
Music datasets can be binned into two variations- 

\smallskip
\noindent \textbf{Symbolic Music Datasets} contain musical scores in formats like MIDI, MusicXML or ABC notation and can sometimes be paired with their corresponding audio. With MIDI datasets being the most popular for example- Lakh MIDI Dataset \cite{raffel2016learning}, Popular examples include Lakh MIDI Dataset \cite{raffel2016learning}, MAESTRO \cite{hawthorne2018enabling}, POP909 \cite{wang2020pop909} and Million-MIDI Dataset (MMD) \cite{zeng2021musicbert}. ABC notation datasets such as Notthingham dataset\footnote{\url{https://ifdo.ca/~seymour/nottingham/nottingham.html}} and Textune \cite{wu2022exploring} have become popular as well for better readability and editing.

\smallskip
\noindent \textbf{Audio Music Datasets} consist of raw audio recordings with additional metadata and are commonly used for tasks such as music generation, classification, and transcription. Notable datasets include MusicCaps \cite{agostinelli2023musiclm}, MusicBench \cite{melechovsky2023mustango} and MuLaMCap \cite{huang2023noise2music}, which provide music clips with descriptive captions and are usually used in tasks like music generation, music captioning and retrieval. GTZAN dataset \cite{sturm2013gtzan} is usually helpful for genre classification, and FMA \cite{defferrard2016fma} for music tagging task.

\subsection{Popular Musical Understanding Tasks}
Figure \ref{fig:ev} illustrates music-related tasks with their corresponding evaluation metrics. Besides generation tasks, computational music research revolves around Music Understanding-related tasks, which include a variety of downstream tasks that are briefly discussed below- 

\smallskip
\noindent \textbf{Music Information Retrieval (MIR)} covers tasks such as key and tempo estimation, genre and style classification, beat detection, chord estimation, instrument identification \cite{raffel2014mir_eval}. MARBLE Benchmark \cite{yuan2023marble} provides a standardized evaluation for 18 such MIR tasks. 

\smallskip
\noindent \textbf{Music Question Answering} involves answering music-related questions based on symbolic or audio input~\citep{deng2023musilingo, liu2024music}. These models are assessed using metrics frequently used such as BLEU, METEOR, and ROUGE-L, and sometimes human evaluation \cite{melechovsky2023mustango} or LLM-based scoring \cite{gardner2023llark}.

\smallskip
\noindent \textbf{Music Captioning} deals with generating lyrics given audio~\citep{gardner2023llark, deng2023musilingo} using joint audio-text representations. Evaluation metrics are fairly similar to Music Question Answering due to the same nature of the output.

\smallskip
\noindent \textbf{Music Retrieval and Recommendation} works with  joint audio-text representations as well to retrieve relevant audio or symbolic music from textual prompts \citep{wu2023clamp, manco2022contrastive} and usually utilizes ranking metrics such as Recall@K, HR@K, and MAP. 

\smallskip
\noindent \textbf{Music Agents} such as MusicAgent \cite{yu2023musicagent}, ComposerX \cite{deng2024composerx}, Loop Copilot\cite{zhang2023loop} are autonomous systems that integrate multiple AI models to perform diverse music-related tasks.

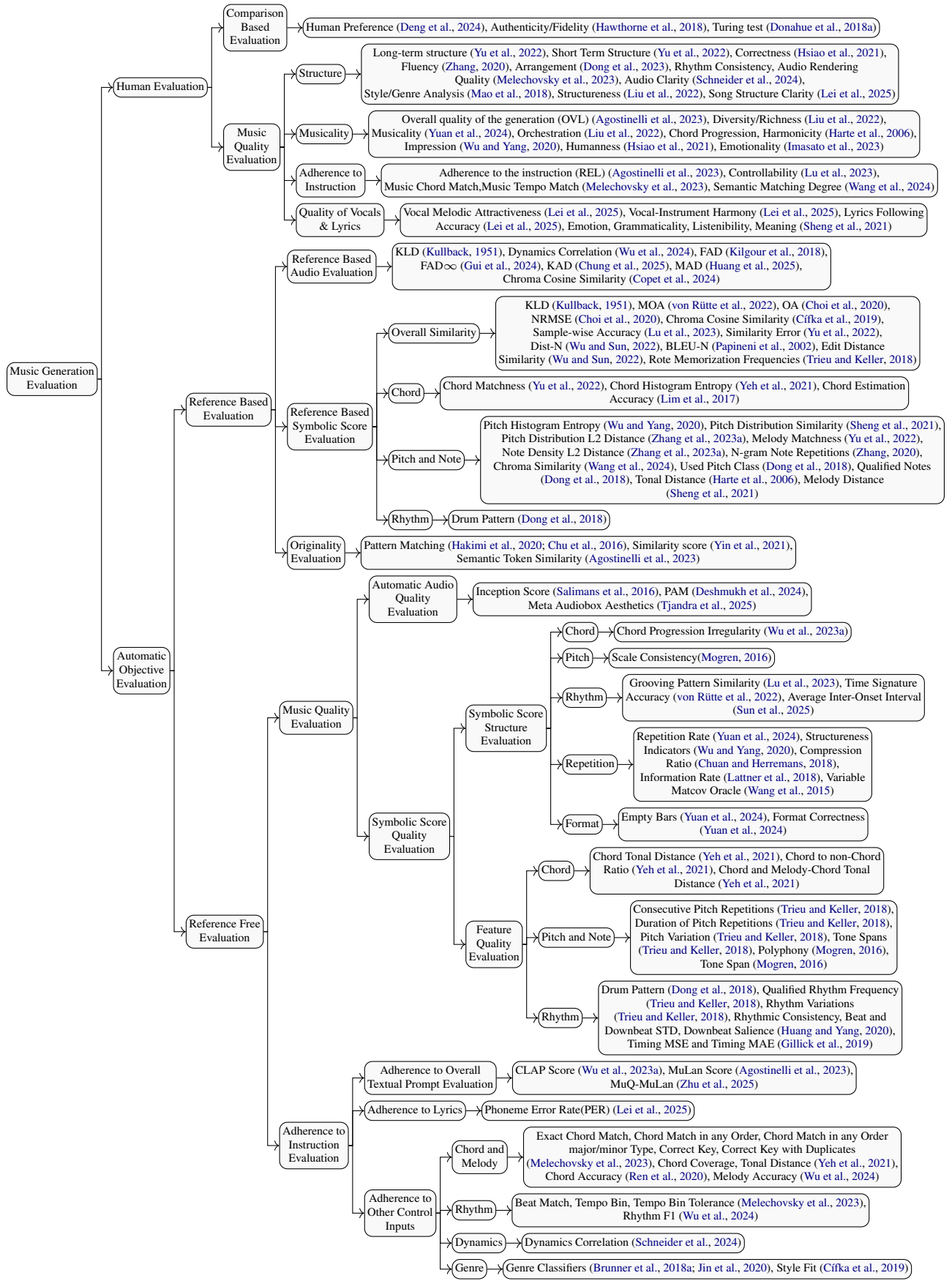
\begin{figure*}[!htb]
\centering
\scriptsize
\begin{forest}
for tree={
font=\tiny,
    draw, rounded corners, fill=gray!5,
    grow'=0,
child anchor=west,
parent anchor=east,
anchor=west,
    calign=center,
    inner sep=1.5pt,          
    l=3pt,                  
    s sep=3pt,              
    align=center,
    edge path={
    \noexpand\path [draw, ->, \forestoption{edge}]
    (!u.east) -- ++(2pt,0) |- (.west) \forestoption{edge label};
},
    % before typesetting nodes={
    %     if n=1
    %         {insert before={[,phantom]}}
    %         {}
    % },
    % fit=band
}
[Music Generation \\Evaluation
   [Human Evaluation
    [Comparison \\Based \\Evaluation  [{ Human Preference \cite{deng2024composerx},
  Authenticity/Fidelity \cite{hawthorne2018enabling},
  Turing test \cite{donahue2018nes}}]
  ]
  [Music \\Quality \\Evaluation
    [{Structure}
      [{ Long-term structure \cite{yu2022museformer},
  Short Term Structure \cite{yu2022museformer},
  Correctness \cite{hsiao2021compound},\\
  Fluency \cite{zhang2020learning},
  Arrangement \cite{dong2023multitrack},
  Rhythm Consistency, Audio Rendering \\Quality \cite{melechovsky2023mustango},
  Audio Clarity \cite{schneider2024mousai},\\
  Style/Genre Analysis \cite{mao2018deepj},
  Structureness \cite{liu2022symphony},
  Song Structure Clarity \cite{lei2025levo}}]]
    [{Musicality} [{Overall quality of the generation (OVL) \cite{agostinelli2023musiclm},
  Diversity/Richness \cite{liu2022symphony},\\
  Musicality \cite{yuanchatmusician},
  Orchestration \cite{liu2022symphony},
  Chord Progression, Harmonicity \cite{harte2006detecting},\\
  Impression \cite{wu2020jazz},
  Humanness \cite{hsiao2021compound},
  Emotionality \cite{imasato2023using}}]]
    [Adherence to \\Instruction[{Adherence to the instruction (REL) \cite{agostinelli2023musiclm},
  Controllability \cite{lu2023musecoco},\\
  Music Chord Match,Music Tempo Match \cite{melechovsky2023mustango},
  Semantic Matching Degree \cite{wang2024melotrans}}]]
    [Quality of Vocals\\\& Lyrics [{Vocal Melodic Attractiveness \cite{lei2025levo},
  Vocal-Instrument Harmony \cite{lei2025levo},
  Lyrics Following \\Accuracy \cite{lei2025levo}, Emotion, Grammaticality, Listenibility, Meaning \cite{sheng2021songmass}}]]
    ]
  ]
  [Automatic \\Objective\\ Evaluation
  [Reference Based \\Evaluation [Reference Based\\
Audio Evaluation[{KLD \cite{kullback1951kullback}, Dynamics Correlation \cite{wu2024music}, FAD \cite{kilgour2018fr},\\
  FAD$\infty$ \cite{gui2024adapting}, KAD \cite{chung2025kad}, MAD \cite{huang2025aligning},\\
  Chroma Cosine Similarity \cite{copet2024simple}}]] [Reference Based \\Symbolic Score\\Evaluation [Overall Similarity [{KLD \cite{kullback1951kullback}, MOA \cite{von2022figaro}, OA \cite{choi2020encoding}, \\NRMSE \cite{choi2020encoding}, Chroma Cosine Similarity \cite{cifka2019supervised},\\ Sample-wise Accuracy \cite{lu2023musecoco}, Similarity Error \cite{yu2022museformer}, \\Dist-N \cite{wu2022exploring}, BLEU-N \cite{papineni2002bleu}, Edit Distance \\Similarity~\cite{wu2022exploring},
Rote Memorization Frequencies~\cite{trieu2018jazzgan}}]] [Chord [{Chord Matchness \cite{yu2022museformer},
  Chord Histogram Entropy \cite{yeh2021automatic},
  Chord Estimation \\Accuracy \cite{lim2017chord}}]][Pitch and Note [{Pitch Histogram Entropy \cite{wu2020jazz}, Pitch Distribution Similarity \cite{sheng2021songmass}, \\Pitch Distribution
  L2 Distance \cite{zhang2023sdmuse}, Melody Matchness \cite{yu2022museformer}, \\Note Density L2 Distance \cite{zhang2023sdmuse}, N-gram Note Repetitions \cite{zhang2020learning}, \\Chroma Similarity \cite{wang2024melotrans}, Used Pitch Class \cite{dong2018musegan}, Qualified Notes \\\cite{dong2018musegan}, 
  Tonal Distance \cite{harte2006detecting}, Melody Distance \\\cite{sheng2021songmass}}]][Rhythm [{Drum Pattern \cite{dong2018musegan}}]]] [Originality \\Evaluation [{Pattern Matching \citep{hakimi2020bebopnet,chu2016song}, Similarity score \cite{yin2021good},\\ Semantic Token Similarity \cite{agostinelli2023musiclm}}]]]
  [Reference Free \\Evaluation [Music Quality \\Evaluation [Automatic Audio\\ Quality \\Evaluation [{Inception Score \cite{salimans2016improved}, PAM \cite{deshmukh2024pam}, \\Meta Audiobox Aesthetics \cite{tjandra2025meta}}]] [Symbolic Score \\Quality \\Evaluation [Symbolic Score\\ Structure \\Evaluation [Chord [ {Chord Progression Irregularity \cite{wu2023tunesformer}}]] [Pitch [{Scale Consistency\cite{mogren2016c}}]] [Rhythm [{Grooving Pattern Similarity \cite{lu2023musecoco}, Time Signature \\Accuracy \cite{von2022figaro}, Average Inter-Onset Interval\\ \cite{sun2025music}}]] [Repetition [{Repetition Rate \cite{yuanchatmusician}, Structureness \\Indicators \cite{wu2020jazz},
  Compression \\Ratio \cite{chuan2018modeling},\\
  Information Rate \cite{lattner2018imposing},
  Variable \\Matcov Oracle \cite{wang2015music}}]] [Format [{Empty Bars \cite{yuanchatmusician}, Format Correctness\\\cite{yuanchatmusician}}]]]
  [Feature \\Quality \\Evaluation [Chord [{Chord Tonal Distance \cite{yeh2021automatic},
  Chord to non-Chord \\Ratio \cite{yeh2021automatic},
  Chord and Melody-Chord Tonal \\Distance \cite{yeh2021automatic}}]] [Pitch and Note [{Consecutive Pitch Repetitions \cite{trieu2018jazzgan}, \\Duration of Pitch Repetitions \cite{trieu2018jazzgan}, \\Pitch Variation \cite{trieu2018jazzgan}, Tone Spans \\\cite{trieu2018jazzgan}, Polyphony \cite{mogren2016c},\\
  Tone Span \cite{mogren2016c}}]] [Rhythm [{Drum Pattern \cite{dong2018musegan}, Qualified Rhythm Frequency\\ \cite{trieu2018jazzgan}, Rhythm Variations \\\cite{trieu2018jazzgan}, Rhythmic Consistency,
   Beat and\\ Downbeat STD, Downbeat Salience \cite{huang2020pop},\\ Timing MSE and Timing MAE \cite{gillick2019learning}}]]]]] [Adherence to \\Instruction \\Evaluation [Adherence to Overall \\Textual Prompt Evaluation [{CLAP Score \cite{wu2023tunesformer}, MuLan Score \cite{agostinelli2023musiclm}, \\MuQ-MuLan \cite{zhu2025muq}}]][Adherence to Lyrics [{Phoneme Error Rate(PER) \cite{lei2025levo}}]][Adherence to \\Other Control \\Inputs [Chord and \\Melody [{Exact Chord Match, Chord Match in any Order, Chord Match in any Order\\ major/minor Type, Correct Key, Correct Key with Duplicates \\\cite{melechovsky2023mustango}, Chord Coverage, Tonal Distance \cite{yeh2021automatic},\\Chord Accuracy \cite{ren2020popmag},
  Melody Accuracy \cite{wu2024music}}]][Rhythm [{Beat Match, Tempo Bin, Tempo Bin Tolerance \cite{melechovsky2023mustango},\\
  Rhythm F1 \cite{wu2024music}}]] [Dynamics [{Dynamics Correlation \cite{schneider2024mousai} }]] [Genre [{Genre Classifiers \citep{brunner2018midi, jin2020style}, Style Fit \cite{cifka2019supervised}}]]]]]]] 
\end{forest}
\caption{Music Generation Evaluation Taxonomy} 
\label{fig:structure}

\end{figure*}

\section{Music Generation Evaluation} 
Generated Music evaluation can be broadly divided into two categories: (1) Subjective evaluation via human judgment and listening tests, and (2) Automatic objective evaluation using computational metrics. This section first reviews objective evaluation methods and their shortcomings, followed by human evaluation methods and lastly discusses ongoing efforts in benchmark development for evaluation. 

\subsection{Automatic Objective Evaluation} \label{audiev}
Automatic objective evaluation encompasses computational methods for assessing generated music. As shown in Figure~\ref{fig:structure}, it includes both reference-based evaluation, which compares generated outputs to ground-truth references across audio and symbolic modalities, and reference-free evaluation, which assesses the generation's quality and structure on its own. 

\subsubsection{Reference Based Evaluation}
Reference-based metrics help assess the extent to which the generation is similar to the target reference. As audio music and symbolic scores are of different modalities (signal and text), their evaluation metrics vary as well and are discussed separately for better clarity. 

\smallskip
\noindent \textbf{Audio Similarity Evaluation:} Most commonly used audio similarity metrics like \textit{KLD (Kullback-Leibler divergence)} and \textit{FAD (Fréchet Audio Distance)} assess how well generated audio matches a target distribution. \textit{KLD} measures the difference between two probability distributions. In music evaluation, these distributions are often derived from the outputs of pretrained audio classifiers (like PANNs \cite{kong2020panns} and PaSST \cite{koutini2021efficient}) or features \cite{chen2024sympac}, allowing KLD to capture a high-level semantic similarity between generated and reference audio sets.

On the other hand, \textit{FAD} \cite{kilgour2018fr} evaluates whether generated audio is plausible and clean by comparing its distribution to a background dataset using embeddings from pretrained audio classifiers and measuring their Fréchet distance. Even though FAD is widely used, its effectiveness depends on the choice of audio classifier \citep{huang2023noise2music, tailleur2024correlation}, reference set quality \citep{gui2024adapting}. It assumes that the audio feature embeddings follow a Gaussian distribution which is often false for real-world audio, whose feature distributions can be complex and non-Gaussian \cite{chung2025kad}. Which should be used as the appropriate audio classifier and reference set for FAD is still debatable \cite{gui2024adapting,lee2024etta, evans2025stable}.

Larger reference sets yield more stable and accurate FAD scores, while small ones cause biased estimates due to poor statistical representation. To correct this, \textit{FAD$\infty$} \cite{gui2024adapting} was proposed which approximates FAD as if computed with an infinite-sized reference set.

To tackle the limitations of FAD, recently newer metrics like \textit{KAD (Kernel Audio Distance)} \cite{chung2025kad} and \textit{MAD (MAUVE Audio Divergence)} \cite{huang2025aligning} metrics were proposed. KAD uses Maximum Mean Discrepancy (MMD) to compare distributions without assuming a Gaussian distribution, making it more reliable with small sample sizes. MAD also avoids the Gaussian assumption which uses self-supervised MERT embeddings and k-means clustering to better capture complex distributions. Both KAD and MAD metrics have shown better correlation with human preferences than FAD. This shows that research efforts are being made to create more perceptually relevant objective evaluation metrics.

\smallskip
\noindent \textbf{Symbolic Score Similarity Evaluation:} Evaluating symbolic music is less standardized than audio evaluation, with many works defining their own metrics and using various symbolic representations. The most common framework, proposed by \citet{yang2020evaluation} which used \textit{Overlapped Area (OA)} and \textit{Kullback-Leibler Divergence (KLD)} to compare pitch and rhythm feature distributions between generated and reference sets. OA and KLD can give us an idea of whether the features from generation are similar to the reference set, or to what extent. While useful, OA computes feature histograms over the entire sequence, failing to account for temporal order. To address this, \textit{Macro Overlapped Area (MOA)} \cite{von2022figaro} was introduced to incorporate temporal order as well. Additionally, less common similarity-based metrics are listed in Figure \ref{fig:structure}.

\smallskip
\noindent \textbf{Originality Evaluation:} A critical task is ensuring that generative models produce novel content rather than simply copying their training data. Earlier methods used pattern matching like n-grams \cite{hakimi2020bebopnet}, longest common subsequence \cite{chu2016song} and cardinality-based similarity scores \cite{yin2021good} to detect overfitting. Recent approaches used exact and approximate semantic token matches \cite{agostinelli2023musiclm} and embedding-based methods, such as LAION-CLAP \cite{wu2023large} embeddings, to identify repeated audio segments, which are then verified through manual listening \cite{evans2024long, evans2025stable}.

\subsubsection{Reference Free Evaluation}
Reference-free metrics address this by assessing two key dimensions: 1) quality of the generation on its own and 2) its adherence to user instructions. 

\smallskip

\noindent \textbf{Music Quality Evaluation:} Music quality evaluation includes both theoretical and perceptual quality evaluation. It involves assessing the structural integrity of the composition based on music theory as well as evaluating whether the music is aesthetically pleasing and emotionally impactful to listeners.

\noindent \textit{\textbf{Automatic Audio Quality Evaluation: }}Even though there is no defined way to quantify audio quality, standalone metrics for perceived audio quality are constantly being developed.
\textit{Inception Score} \cite{salimans2016improved} is used to assess quality and diversity but can be misleading if a model overfits on its training data \cite{donahue2018adversarial}. 

\textit{PAM} \cite{deshmukh2024pam} assesses overall audio quality without a reference by using an audio-language model to detect distortions and artifacts by comparing an audio sample against contrasting text prompts ("clear sound" vs. "noisy sound"). \textit{Audiobox Aesthetics} \cite{tjandra2025meta} is a domain-agnostic model trained on 97,000 annotated clips to predict four distinct and interpretable aesthetic dimensions- Production Quality, Production Complexity, Content Enjoyment and Content Usefulness. The latest trend involves training aesthetic predictors \citep{yao2025songeval} directly on large-scale human preference datasets \citep{huang2025aligning, liu2025musiceval, yao2025songeval}.
Human preference datasets mainly contain generative songs that are annotated with human preference ratings (details of the datasets are discussed in \ref{benchmark}). Even though newer works \citep{yuan2025yue, zhang2025aesthetics, gong2025ace} have quickly started to adapt Audiobox Aesthetics in their evaluation, \cite{yao2025songeval} showed that models trained on their human preference dataset, SongEval outperform Audiobox Aesthetics in predicting human-perceived musical quality.

\noindent \textit{\textbf{Symbolic Score Quality Evaluation:}} Symbolic Score Quality Evaluation remains less advanced compared to audio quality evaluation as well. It typically involves manual or rule-based analysis to assess the structural correctness of the score and the quality of the features. 

\noindent \textit{\textbf{A) Symbolic Score Structure Evaluation:}}
These metrics can be utilized to check if the generation is maintaining a proper structure and adhering to the music theory or not. Checking for irregularity in chords \cite{wu2023tunesformer}, rhythmic consistency \cite{lu2023musecoco}, and scale consistency \cite{mogren2016c} are some ways to check for feature-wise structures in generations, but the use of these metrics is not standardized. \textit{Empty Bars (EB)} (ratio of empty bars) and \textit{Format Correctness Evaluation} \cite{yuanchatmusician} are used for calculating syntactical accuracy. Some works \citep{yuanchatmusician, wu2020jazz, chuan2018modeling, lattner2018imposing, chen2019effect} checked for repeating patterns in the generated score, as it can indicate music-like structure.

\noindent \textbf{\textit{B) Feature Quality Evaluation:}} These metrics are feature heavy and may provide some insight into the quality of specific musical features used, however, are no way sufficient to quantify the overall music quality. There is Figure \ref{fig:structure} lists the metrics used for checking the quality of \textit{Chords}, \textit{Pitch and Note} and \textit{Rhythm} respectively. Visualizing tools such as- \textit{Spectrogram of generated waveforms} \cite{zhu2023ernie}, \textit{Constant-Q Transform spectrograms} \cite{engel2017neural}, \textit{Pianorolls} \cite{dong2018musegan}, \textit{Keyscapes} \cite{lattner2018imposing}, \textit{Fitness scape plots}\cite{muller2012scape} can be utilized to assess feature quality visually.

\noindent \textbf{Adherence to Instruction Evaluation: }Adherence to Instruction Evaluation measures how well generated music aligns with input directives, which can be textual prompts or structured controls like lyrics, chords or style, ensuring the output faithfully reflects the intended guidance.

\noindent \textbf{\textit{Adherence to Textual Prompts Evaluation: }} For text-to-music models, adherence to textual prompts is typically measured by computing the cosine similarity between the text embedding of the prompt and the audio embedding of the generation. While CLAP Score \cite{huang2023make, evans2024long} is common where embeddings are derived from CLAP \citep{elizalde2023clap, wu2023large} models, it is a non-music specific model. Other alternatives like MuLan embeddings, MuQ-MuLan \cite{zhu2025muq} and CLAMP 3 model \cite{wu2025clamp} showed better performance due to being trained on more music-aware tasks and larger datasets \citep{agostinelli2023musiclm, gong2025ace,yuan2025yue}.

\noindent \textbf{\textit{Adherence to Lyrics: }}\textit{Phoneme Error Rate(PER)} is used to check how well the given lyric aligns in the generated audio. PER is calculated by extracting the vocal track and passing that to a lyrics recognition model \cite{lei2025levo}. \citet{sheng2021songmass} evaluated alignment accuracy of the melody and lyrics to ensure structural consistency.

\noindent \textbf{\textit{Adherence to Other Control Inputs:}} Control inputs for symbolic music generation other than textual descriptions can affect the selection of evaluation metrics. Some works \citep{wu2024music, melechovsky2023mustango, yeh2021automatic, ren2020popmag} evaluated fine-grained feature control ability of their models by using few feature specific metrics listed in figure \ref{fig:structure}, but use of these metrics are less common in literature. style or genre adherence is often evaluated using a dedicated classifier \citep{brunner2018symbolic, jin2020style}. Since classifier scores only indicate the presence of some distinguishing features rather than true stylistic conformity, \citet{cifka2019supervised} proposed a more interpretable style fit metric to evaluate stylistic alignment. In emotion-controlled generation, discriminator models have been used to classify whether a generated piece belongs to the intended emotional category \cite{imasato2023using}.

Appendix \ref{def} lists some of metric definitions and appendix \ref{tool} mentions currently available toolkits used for evaluation, which were skipped over due to space shortage.

\subsection{Human Evaluation} \label{humev}
Since there is still no clear method to assess creativity and musical quality, most music generation evaluations rely on human judgment for validation. Human evaluation involves designing appropriate listening experiments with logically useful assessment criteria involving appropriate candidates and environment to qualitatively evaluate generated music. 
In \textbf{Comparison based} listening tests, listeners are often asked to compare two or more samples. This can be called a Turing Test, where the goal is to distinguish between human-composed and AI-generated music \citep{lee2022commu, donahue2018nes, donahue2019lakhnes}, or a preference test asking which sample is of higher quality \citep{deng2024composerx, hawthorne2018enabling}.

Other than comparison, participants rate generated music on one or more criteria, typically using a Likert scale \cite{huang2023make} or by providing a Mean Opinion Score (MOS) \cite{liu2025musiceval}. Assessment criteria are much less standardized as works \citep{melechovsky2023mustango, jin2020style} usually define their own assessment criteria and can be broadly categorized into these evaluation aspects- 

\begin{itemize}
    \item \textbf{Musical structure according to music theory} assesses how well the audio follows logical and theory-aligned musical organization.
    
    \item \textbf{Music quality} captures aspects like creativity, harmonic richness, and emotional impact.
    
    \item \textbf{Adherence to instruction} measures how accurately the output reflects the given prompt.
    
    \item \textbf{Quality of vocals} evaluates the attractiveness and harmonic integration of vocals in the audio.
\end{itemize}

Figure \ref{fig:structure} has the assessments listed typically used in human evaluation and Appendix \ref{human} discusses their definition. A listening test design can be task-specific as well, for example- \cite{jin2020style} conducted a listening test to evaluate classical music generation and defined own assessments criteria with respect to the characteristics of only classical songs. \cite{suzuki2023comparative} used OpenAI's ChatGPT and Google's Bard to assess the generated music's atmosphere and genre as well as their human evaluation counterpart on these exact metrics. Hypothesis tests such as Kruskal-Wallis H test, Wilcoxon signed-rank test, t-tests are done to validate the statistical significance of the human ratings \citep{donahue2019lakhnes,hawthorne2018enabling}.

\subsection{Benchmarks} \label{benchmark}
MusicCaps \cite{agostinelli2023musiclm},  MusicBench \cite{melechovsky2023mustango} and Song Describer Dataset \cite{manco2023song} are often used to evaluation text-to-audio music (TTM) generation models\footnote{\url{https://paperswithcode.com/task/text-to-music-generation\#datasets}} \citep{evans2024long, evans2025stable}.
Ziqi-Eval's music generation question set \cite{li2024music} offers 184 multiple-choice and 200 five-shot questions to test LLMs on melody continuation, technically assessing music understanding rather than generation capabilities. Several human preference datasets have been proposed- MusicPrefs \cite{huang2025aligning} with 183,000 clips and crowdsourced pairwise ratings for fidelity and musicality. Dynamo Music Aesthetics (DMA) \cite{bai2025dragon} includes 800 prompts, 1,676 pieces (15.97 hours) and 2,301 detailed 1–5 ratings from 63 raters. MusicEval \cite{liu2025musiceval} contains 2,748 clips from 31 TTM models with over 13,000 expert ratings for musical impression and text alignment.
SongEval \cite{yao2025songeval} is a large-scale benchmark of 2,399 songs (140+ hours), rated by 16 professionals across five dimensions: coherence, memorability, naturalness, clarity and musicality.

\begin{figure*}[!htb]  
    \centering
   \includegraphics[width=13cm, height=7cm]{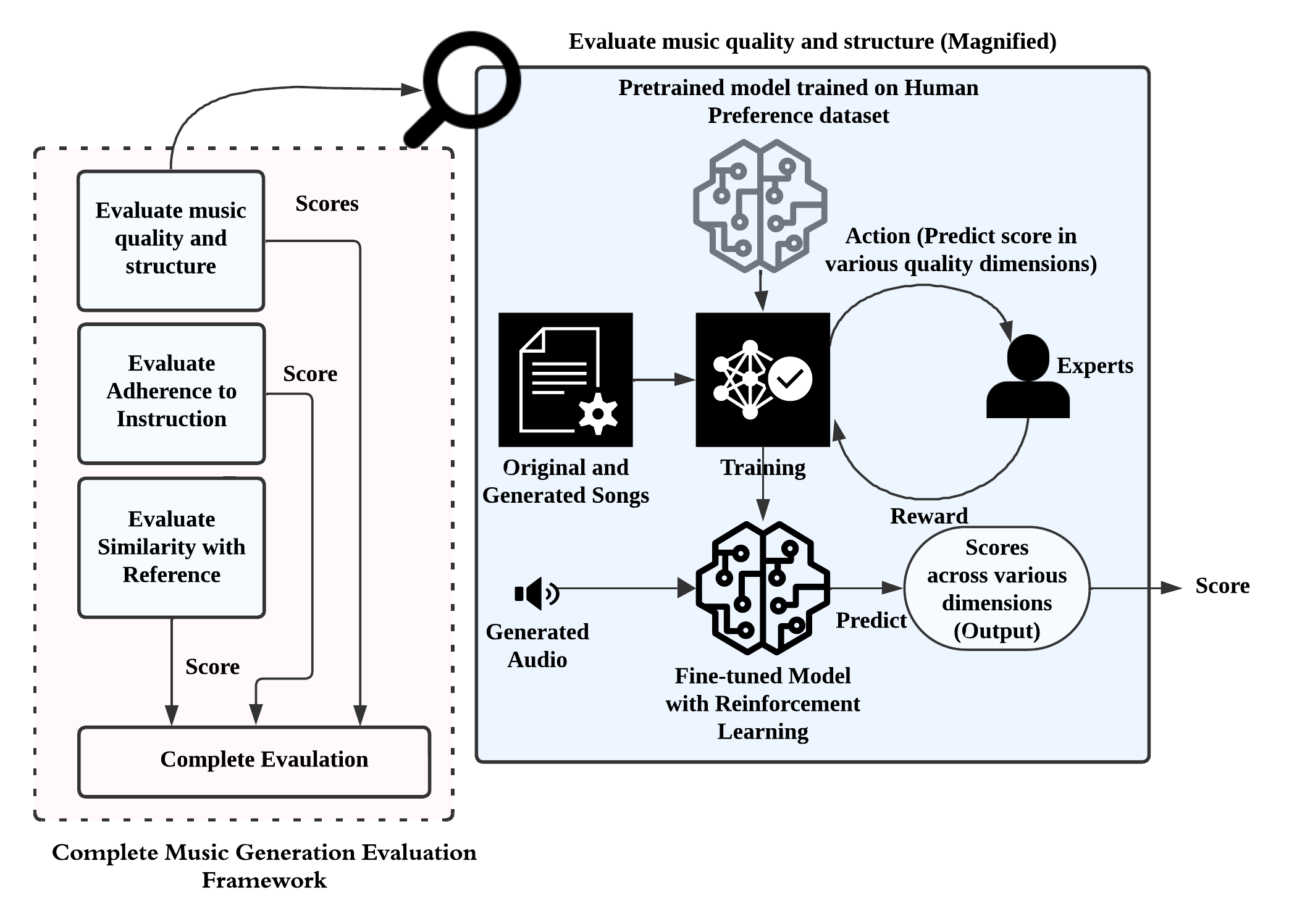}  
     \caption{A reliable Music Quality Scorer Model can elevate the current music generation evaluation scenario.}
    \label{fig:framework}
    \vspace{-2mm}
\end{figure*} 

\section{A Critical Review}
In this section, we present some critical analyses of the current music generation evaluation metrics, followed by identifying research gaps and pathways for future research to overcome them.
\subsection{A Critical Analysis of Objective Metrics}
\noindent \textbf{Limitations of Similarity-Based Metrics :} High scores on similarity-based metrics do not guarantee high-quality or musically meaningful compositions. Similarity with target distribution simply means generated scores show similar characteristics as the reference set, but no way quantifies if the piece itself is a good sounding piece or a distuned boring sounding piece. Unless it is a controlled generation, syntactical similarity metrics like BLEU, Average Sample-wise Accuracy and Chord Matchness can easily seem useless for the same reason. Only assessing the similarity with the reference leads to an incomplete evaluation and should be accompanied with reference free music quality evaluation. 

\noindent \textbf{Lack of Interpretation :} \citet{yuan2025yue} showed that many widely used objective metrics, such as CLAP-score, FAD, and KLD, often align poorly with human preferences, which makes the conclusions of prior studies that rely on these measures questionable. A core issue is that these metrics lack clear interpretability. For example, metrics like OA and KLD score are considered as the higher the better, but have no meaningful threshold or guidance for balancing similarity and originality. Similarly, Chord Progression Irregularity \cite{wu2023tunesformer} measures the percentage of unique chord trigrams, where lower values suggest greater stability yet extremely low values can be interpreted as a boring sequence as well. While these scores can rank models and indicate feature quality, better scoring outputs may not correspond to better sounding to listeners. Overall, objective metrics alone can’t reliably evaluate musical quality and risk misrepresenting what truly sounds good without human evaluation.

\noindent \textbf{Lack of Cross-cultural Consideration :} A significant limitation in music generation evaluation arises from cross-cultural biases in datasets, benchmarks, and evaluation methods. \citet{mehta2025music} quantified the severe Western-centric bias in 152 musical dataset proposing papers, finding only 5.7\% of music comes from non-Western genres, including South Asian, Middle Eastern, Oceanian, Central Asian, Latin American and African music combined. Models trained on these datasets struggle to generate low-resource genres, while evaluation metrics tailored to Western styles may fail to assess diverse musical characteristics and lack suitable training data. For example, FAD’s reliance on the choice of reference set and audio embeddings raises concerns that it may favor only well-resourced genres. Another example can be, in Pitch Histogram Entropy, a high entropy suggests unstable tonality and pitch classes are more scattered which may favor straightforward genres like pop but is ill-suited for evaluating microtonal, polyrhythmic or improvisational music from low resource traditions. Similarly, corpus-based evaluations favor well-documented styles while overlooking culturally unique ones. \citet{wilson2025short} further highlighted limited transparency, with few models disclosing training data or generation methods, hindering efforts to address these biases.

\noindent \textbf{Lack of Standardization :} While feature-specific metrics can be useful for analyzing individual systems, they often fail to generalize, with many researchers adapting their own evaluation criteria. The resulting overwhelming number of specialized metrics make it difficult to determine which are truly effective, hindering clear assessment and comparison of different generative models' strengths and weaknesses.

\noindent \textbf{Limitations of Music Quality/Aesthetic Predictors :} For efficiency and growing need of large-scale evaluation, recent works are shifting towards automatic music quality evaluation using aesthetic predictors like Audiobox Aesthetics \cite{tjandra2025meta}. Unfortunately, \citet{zhang2025aesthetics} highlighted that human preference datasets often misalign with these independently trained aesthetic predictors. This indicates that human preference is not a single, consistent concept as human perception of creativity is subjective and shaped by geography, history, and culture \cite{lubart1999creativity}. Different evaluation methods, even if both are based on human feedback, can lead to contradictory conclusions about music quality, raising concerns about their reliability and generalizability. \citet{zhang2025aesthetics} further showed aesthetic predictors favor certain content, with tracks featuring "punchy kick" or "synth".

\noindent \textbf{Limitations of Human Preference Datasets:} Human preference datasets can introduce bias as they are constructed with generative audios from current TTM models which often fail with low-resource genres as well. Furthermore most of the human preference datasets rely solely on overall impression \cite{huang2025aligning, liu2025musiceval} or preference \cite{bai2025dragon} which is insufficient for modeling human perception of musical creativity. We have to further break down the judgment of creativity into several equal dimensions and employ experts to rate audios across these dimensions. A simple example of why this works is, despite individual taste differences, expert food critics evaluate dishes across equally important dimensions like flavor, texture, presentation, originality, execution, and overall impression. Among the human preference datasets, SongEval \cite{yao2025songeval} broke music quality evaluation into multiple dimensions, but further analysis with music experts is needed to ensure the dimensions are enough to cover all the aspects of quality evaluation.

\noindent \textbf{Limitations of Symbolic Music Evaluation : }Symbolic music generation evaluation improvement is lagging behind audio music evaluation in both standardization and depth of analysis, largely because symbolic representations lack direct perceptual grounding. While audio evaluation heading towards using perceptual and embedding-based metrics that can align well with human perception, symbolic evaluation often relies on simplistic feature based measures that might miss important aspects of music quality, creativity, and correlation with human perception. Furthermore, symbolic evaluation lacks standard benchmarks, representations and validated features, making it hard to compare models or ensure metrics generalize across styles.

\subsection{A Critical Analysis of Human Evaluation}
\noindent \textbf{Sensitivity to Participant Background :}
Designing a listening test can be challenging as they are highly sensitive to factors like- variation in participant expertise and uneven participant group sizes, followed by biases due to age, education, cultural exposure, cognitive traits \citep{ferreira2023generating, yang2020evaluation}. The chances of these biases increase when participants come from a single background, limiting generalizability. \citet{ferreira2023generating} conducted a blind listening test with 117 participants from diverse backgrounds to evaluate their ability to distinguish between AI-generated and human-composed music. Results showed that frequent classical music listeners, musicians and individuals with high self-assessed musical sensitivity were significantly more accurate in identifying the source, highlighting the need to appoint raters with appropriate musical background and perceptual skill.

\noindent \textbf{Experimental Design Challenges :} Aside from participant expertise, design of a listening test is not standardized as well with factors to consider like- sample selection, environment setting of the listening test and phrasing of the surveys. Environment variations, confusing phrasing of the surveys and small sample sizes reduce statistical reliability. For example, in their listening test, \citet{schneider2024mousai} defined musicality as how much the given sound is melodiousness and harmoniousness, whereas \citet{yuanchatmusician} defined musicality based on two aspects- the overall consistency of the music in terms of melodic patterns and chord progressions etc. and the presence of a clear structural development with respect to features.
With works designing their own listening test criteria and the high cost of large-scale studies is a big setback for standardized, cross-model comparisons \cite{yang2020evaluation}. 

\subsection{Summary of Major Limitations}
Among the challenges in music generation evaluation discussed in previous section, several stand out as particularly critical. The lack of interpretability and reliability of objective metrics undermines the evaluation's ability to draw meaningful conclusions, as widely used measures often misalign with human perception and lack clear thresholds for quality. The lack of cross-cultural consideration introduces severe biases by favoring Western music traditions in datasets and evaluation methods. The lack of standardization in evaluation methodologies make cross-model comparisons difficult as well. Finally, limitations in designing a listening test for human evaluation weaken the validity of listener studies intended to capture subjective musical quality. Unfortunately, these limitations question the credibility and inclusiveness of music generation evaluation methods which calls for the urgent need for more interpretable, culturally aware standardized evaluation frameworks.

\subsection{Research Gaps}
This section identifies three open research questions in music generation evaluation paradigm, each illustrating a distinct category of research gap. First, despite extensive study, the question \textit{“How to model and evaluate creativity in music? Does modeling human perception automatically model creativity?”} remains unresolved, as existing methods struggle to deliver robust or generalizable solutions for capturing the subjective nature of creativity. Second, the question \textit{“Can the existing evaluation methods cater to underrepresented genres?”} is currently understudied, requiring better evaluation methods for underrepresented genres. Third, \textit{“How can future efforts in music evaluation develop robust methodologies that effectively integrate computational analysis with listener perception studies and task-specific benchmarks?”} represents an area yet to be systematically explored with the joint efforts of music experts and cognitive scientists to design a comprehensive evaluation frameworks.

\subsection{Opportunities and Future Directions}
We think there should be 3 components for a comprehensive music generation evaluation framework: 1) evaluating music quality and structure, 2) adherence to instruction, and 3) evaluating similarity with reference, respectively (figure \ref{fig:framework}). Future efforts in music evaluation should focus on developing more robust and generalized evaluation methodologies that integrate computational analysis with listener perception studies, cross-cultural considerations and task-specific benchmarks for these 3 components. We welcome the ongoing efforts to emulate human perception of music through automatic aesthetic predictors and human preference datasets for large scale evaluation, but significant research effort is needed to break down the concept of music quality and structure into smaller, definable dimensions whose scores can jointly give us an interpretable way to quantify music quality, rather than only depending on confusing terms such as “overall quality”. We further propose a possible automatic music quality and structure evaluation framework that incorporates the idea of human-in-the-loop training and Reinforcement Learning \cite{kaelbling1996reinforcement} to rank the subjective quality of a generated music according to human perception across scientifically defined dimensions. Starting with a pretrained model on such a human preference dataset, the model will receive original songs as well as generated songs as inputs and predict scores across pre-defined dimensions of music quality. These predictions can be compared with expert ratings to compute a reward to fine-tune the model. Through this feedback loop, the model can learn to align its predictions more closely with human perception. The catch is to have experts from various cultural backgrounds and use original songs specially for low-resource genres to make the aesthetic predictor model less biased and more generalizable. 

\section{Conclusion}
Evaluation for music generation is still a complex challenge due to the inherent subjectivity of music as we are yet to discover how to quantify human perception of creativity. With the recent efforts to model human perception with automatic aesthetic predictors, it is at a very early stage where further research with cognitive scientists and music experts is absolutely necessary to determine modular interpretable evaluation dimensions that would quantify overall quality of a music piece. Furthermore, it is equally necessary to acknowledge and address the biases and lack of interpretation present in current music generation models and evaluation methodologies to make music generation more generalizable to the global music community. 
% \input{latex/limitation}
% \section*{Acknowledgments}
\bibliography{acl2023}
\bibliographystyle{acl_natbib}

\appendix
\clearpage
\appendix
\section{Human Evaluation} \label{human}
\subsection{Musical Structure according to Music Theory }
\textbf{Structureness:} If the music is structured nicely or not \cite{liu2022symphony}. More fine-grained structural aspects were used by \cite{yu2022museformer}. \textbf{Short-term structure:} Whether the generated score is showing good structures, good repetitions and reasonable development within a close range. \textbf{Long-term structure:} Whether the generated score is showing good structures, song-level repetitions and long distance connections within a broader range.

\noindent \textbf{Correctness: } Does the listener perceive any absence of composing or playing mistakes \cite{hsiao2021compound}.

\noindent \textbf{Fluency: } If the generated music sounds fluent or not \cite{zhang2020learning}.

\noindent \textbf{Arrangement: }Are the instruments used reasonably and arranged properly? \cite{dong2023multitrack}.

\noindent \textbf{Rhythm consistency:} Is the rhythm staying constant throughout the music? \cite{melechovsky2023mustango}

\noindent \textbf{Audio Rendering Quality: } To check the audio rendering quality for generated audio \cite{melechovsky2023mustango}.

\noindent \textbf{Audio clarity:} How close the quality is to a walkie-talkie (worst) or a high-quality studio sound system (best) \cite{schneider2024mousai}.

\noindent \textbf{Style/Genre Analysis:} If the generated music can be classified to any genre \cite{mao2018deepj}.

\noindent \textbf{Coherence:} Do the music lines sound coherent or not? \cite{liu2022symphony}

\noindent \textbf{Orchestration:} Is the score nicely orchestrated \cite{liu2022symphony}

\subsection{Music Quality}
\noindent \textbf{Rhythm: } If the note durations and pauses of the melody sound natural or not \cite{sheng2021songmass}.

\noindent \textbf{Diversity/Richness:} How diverse and interesting is the generated musical score \cite{liu2022symphony}, \cite{wu2020jazz}.

\noindent \textbf{Impression: } Does the listener remember some part of the melody \cite{wu2020jazz}.

\noindent \textbf{Humanness:} Does the piece resemble expressive human performances? \cite{hsiao2021compound}

\noindent \textbf{Chord Progression:} Assesses how coherent, pleasant, or reasonable the progression is on its own, independent of melody \cite{harte2006detecting}.

\noindent \textbf{Harmonicity:} Measures how well the progression harmonizes with a given melody \cite{harte2006detecting}.

\noindent \textbf{Interestingness:} Evaluates how exciting, unexpected, or positively stimulating the progression sounds. These three criteria were used to assess models for melody harmonization task.

\noindent \textbf{Emotionality:} How the emotion is perceived in the generated score. Evaluators were asked to place the perceived emotion of each piece on Russell's circumplex model of affect \cite{imasato2023using}.

\noindent \textbf{Innovativeness:} Originality in style, timbre, and structural elements

\subsection{Adherance to Instrution}
\textbf{Semantic Matching Degree (SMD):} How well the generated music matches the expressiveness described by the input text \cite{wang2024melotrans}.

\noindent \textbf{Controllability:} How well the score is adhering to the musical attributes specified in given prompt/text description \cite{lu2023musecoco}.

\noindent \textbf{Music Chord Match} and \textbf{Music Chord Match:} measures to what extent the chords and tempo from the generated music match the text prompt respectively \cite{melechovsky2023mustango}.

\noindent To evaluate generated lyrics from given melody and vice versa these metrics can be utilized-

\noindent \textbf{Listenibility: } Does the lyric sound natural with the melody? \cite{sheng2021songmass}

\noindent \textbf{Grammaticality:} Is the lyric grammaticaly correct? \cite{sheng2021songmass}

\noindent \textbf{Meaning: } If the lyrics seem meaningful or not \cite{sheng2021songmass}.

\noindent \textbf{Emotion: } If the emotion of the melody aligns with the lyrics or not \cite{sheng2021songmass}.

\section{Evaluation Metrics Definition and Behaviour} \label{def}
\noindent \textbf{Pitch and Rhythm Variations }\cite{trieu2018jazzgan} measures the number of unique pitches and note durations within a sequence respetively.

\noindent \textbf{Used Pitch Class (UPC)}\cite{dong2018musegan} is number of used pitch classes per bar.

\noindent \textbf{Qualified Note (QN)}\cite{dong2018musegan} is the proportion of notes that are at least three time steps long (equivalent to a 32nd note or longer). This metric indicates whether the music is too fragmented, with a higher QN suggesting smoother, continuous music.

\noindent \textbf{Drum Pattern (DP)}\cite{dong2018musegan} is the ratio of notes in 8 or 16-beat patterns. The authors suggested that Rock songs frequently use 4/4 beat pattern.

\noindent \textbf{Tonal Distance (TD)}\cite{harte2006detecting} measures harmonicity between two sequences, where a higher tonal distance (TD) indicates weaker harmonic alignment between them.

\noindent \textbf{Qualified Rhythm Frequency}\cite{trieu2018jazzgan} extends \cite{dong2018musegan}'s Qualified Note metric (which excluded notes shorter than a 32nd note) by measuring how often note durations match standard values (1, 1/2, 1/4, 1/8, 1/16) including dotted, triplet, and tied forms.

\noindent \textbf{Consecutive Pitch Repetitions (CPR)}\cite{trieu2018jazzgan} measures the frequency of occurrences of some number of consecutive pitch repetitions. A high CPR represents monotonous repetition in generated music.

\noindent \textbf{Durations of Pitch Repetitions (DPR)}\cite{trieu2018jazzgan} measures how often a pitch is repeated for at least some total duration, helping to detect long repetitions.

\noindent \textbf{ Tone Spans (TS)}\cite{trieu2018jazzgan} counts how often pitch changes exceed a tone distance d (in half-steps).

\noindent \textbf{Polyphony}\cite{mogren2016c} measures the frequency of two tones playing simultaneously.

\noindent \textbf{Melody Distance} \cite{sheng2021songmass} computed Melody distance by normalizing note pitches (subtracting the mean) and comparing generated and ground-truth pitch time series of varying lengths using dynamic time warping.

\noindent \textbf{Information Rate (IR)}\cite{lattner2018imposing} is calculated as the mutual information between present and past observations, where high values indicate structured self-similarity in the generated music. The IR metric is estimated using a first-order Markov Chain, contrasting prior entropy with conditional entropy, making it suitable for assessing the repetition structure of musical sequences.

\noindent \textbf{Rhythmic Consistency} \cite{huang2020pop} measured the Rhythmic Consistency of their generated Pop music compositions by generating 1,000 sequences and analyzing their beats and downbeats using an RNN-DBN model.

\noindent \textbf{Chord Coverage } \cite{yeh2021automatic} counts how many different chord types appear in a chord sequence by checking non-zero values in the chord histogram. It helps assess whether the model is generating a wide variety of chords or sticking to a limited set.

\noindent \textbf{Chord Tonal Distance (CTD)} \cite{yeh2021automatic} measures the average tonal distance \cite{harte2006detecting} between each pair of adjacent chords in a sequence. A higher CTD means there are more abrupt changes in the chord progression. 

\noindent \textbf{Chord Tone to Non-Chord Tone Ratio (CTnCTR)} \cite{yeh2021automatic} is the ratio of notes that match the underlying chord (chord tones) to those that don’t (non-chord tones). A higher CTnCTR indicates that most notes fit well with the chords.

\noindent \textbf{Pitch consonance score (PCS)} \cite{yeh2021automatic} measures how well melody notes fit with the chords. The average consonance score across 16th-note windows  is calculated by checking the musical interval between the melody note and the chord notes.

\noindent Extending the idea of tonal distance, \textbf{Melody-chord tonal distance (MCTD)} \cite{yeh2021automatic} measures the average tonal distance (each distance weighted by the duration of the respective melody note) between each melody note and its corresponding chord label throughout a melody sequence. CC, CTD, CTnCTR, PCS, MCTD help determine how smooth or abrupt chord changes are in the sequence and how well the whole piece harmonizes together.

\noindent \textbf{Alignment accuracy} \cite{sheng2021songmass} measures if the generated melody is accurately aligned with the lyrics by comparing the number of generated tokens with the ground truth.

\noindent \textbf{Variant Proportion (VP$_i$)} \cite{wang2024melotrans} calculates the proportion of the i-th type of variant whether the distribution of variant type is reasonable.

\noindent \textbf{Variant Distance (VD)} \cite{wang2024melotrans} calculates the average length (in beats) to assess whether the model generates variants correctly.

\noindent \textbf{Similarity Error} \cite{yu2022museformer} evaluates pitch and rhythm by creating note sets per bar (including pitch, duration, and onset), then computing mean intersection-over-union (IoU) similarity across bar pairs. The final score is the difference in mean IoUs between original and generated pieces.

\noindent \textbf{Melody Matchness} \cite{yu2022museformer} calculated Melody Matchness in REMI format by finding the bar wise longest common subsequence between the ground truth and generated piano melodies. Two notes are considered a match if they have the same pitch and their onset times are within an eighth note of each other.

\noindent \textbf{Pitch Class Histogram Entropy } \cite{wu2020jazz} To calculate pitch histogram entropy, we can create a 12-dimensional pitch class histogram with the notes that appear in a certain period of the music score and calculate the entropy of that histogram. 
\begin{equation}
H = - \sum_{i=1}^{12} p_i \log_2 p_i
\label{eq:pce}
\end{equation} 
where $H$ is the Pitch Class Entropy.
$p_i$ is the probability of the i-th pitch class (C, C\#, D, ..., B) occurring in a piece. 
Low entropy indicates clear tonality with dominant pitch classes, while high entropy suggests unstable, scattered tonality. Chord Histogram Entropy \cite{yeh2021automatic} applies the same idea to chords.

\noindent \textbf{Pitch and Duration Distribution Similarity} \cite{sheng2021songmass} is the measurement of how similar the pitch and durations distributions are of the generated music and ground truth. First pitch and duration frequency histogram is computed and the similarity is measured by the average overlapped area between the two histograms.

 \noindent \textbf{Chroma similarity} \cite{wang2024melotrans} For symbolic music, particularly in REMI representation, Chroma similarity or $sim_{chr}$, 
measures the closeness of two bars of the generated and reference scores in tone via:
\begin{equation}
   sim_{chr}(ra, rb)=100 <r^a, r^b>/||ra|| ||rb|| 
\end{equation} 
 where \(<.,.>\) denotes dot-product and $r \in Z^{12}$ is the chroma vector representing the number of onsets for each of the 12 pitch classes.

 \noindent \textbf{Macro Overlapped Area (MOA)}\cite{von2022figaro} Let x and y denote two musical sequences and let $b^{(x)}_i$ and $b^{(y)}_i$ denoting their i-th bars. Feature overlap is computed using the Gaussian distributions of a chosen feature, with overlap given by $overlap( b^{(x)}_i, b^{(y}_i)$. Then the macro OA (MOA) between x and y is- 
\begin{equation}
  MOA(x,y) = 1/N\sum_{i=1}^{N} overlap( b^{(x)}_i, b^{(y}_i)  
\end{equation}
 \noindent \textbf{Chord matchness} \cite{yu2022museformer} measured Chord matchness of the generated piano segment and the target chord in the lead sheet by computing the cosine similarity between their respective chroma vectors.

\noindent \textbf{Average Sample-wise Accuracy (ASA)}\cite{lu2023musecoco} is computed by first measuring the proportion of correctly predicted attributes for each sample, then averaging these values across the entire test set.

\noindent \textbf{Dynamics correlation} \cite{wu2024music} measures how well a generated audio score matches the dynamic variations (smoothed frame wise loudness) of a reference performance by calculating Pearson’s correlation.

\noindent \textbf{Grooving Pattern Similarity} \cite{wu2023tunesformer} between a pair of grooving patterns $\vec{g}^{\,a}, \vec{g}^{\,b}$ is calculated by- 
\begin{equation}
\mathcal{GS}(\vec{g}^{\,a}, \vec{g}^{\,b}) = 1 - \frac{1}{Q} \sum_{i=0}^{Q-1} \text{XOR}(g_i^a, g_i^b),
\end{equation}
where $Q$ is the dimensionality.

\noindent \textbf{Structureness Indicators} \cite{wu2020jazz} quantifies musical repetition by analyzing a fitness scape plot, a matrix $S \in \mathbb{R}^{N \times N}$ where each entry $S_{ij} \in [0, 1]$ reflects the degree of repetition for a segment of duration $i$ centered at time $j$. To capture the most prominent structural repetition within a specific time range $[l, u]$, the indicator is defined as $\mathrm{SI}_{u}^{l}(S) = \max_{\substack{l \leq i \leq u \\ 1 \leq j \leq N}} S_{ij}$.

\noindent \textbf{Chord Accuracy} \cite{ren2020popmag} checks if the conditional chord sequence matches the chords of the generated score by calculating-
\begin{equation}
\text{CA} = \frac{1}{N_{\text{tracks}} \cdot N_{\text{chords}}} 
\sum_{i=1}^{N_{\text{tracks}}} \sum_{j=1}^{N_{\text{chords}}} 
\mathbb{I}\{ C_{i,j} = \hat{C}_{i,j} \},
\end{equation}
where $N_{\text{tracks}}$ and $N_{\text{chords}}$ are the number of tracks and chords per track respectively.

\section{Evaluation Toolkits} \label{tool} Several open-source toolkits are available to facilitate evaluation. For symbolic music- MGEval \cite{yang2020evaluation}, MusPy \cite{dong2020muspy}, Music21 \cite{cuthbert2010music21} and JSymbolic \cite{mckay2018jsymbolic} for feature extraction, dataset management, and visualization tools. They support analyzing different features for both absolute and comparative evaluation. For audio music- FAD toolkit\footnote{\url{https://github.com/microsoft/fadtk}}, Stability AI’s code\footnote{\url{https://github.com/Stability-AI/stable-audio-metrics}} for $\text{FD}{openl3}$, $\text{KLD}{passt}$ and $\text{CLAP}_{score}$\cite{evans2024long} calculation, and Meta's Audiobox Aesthetics \footnote{\url{https://github.com/facebookresearch/audiobox-aesthetics}}.

\end{document}